# Radio Frequency Electrical Transduction of Graphene Mechanical Resonators

Yuehang Xu[*,1,2], Changyao Chen[*,1], Vikram V. Deshpande[*,3], Frank A. DiRenno[1], Alexander Gondarenko[1], David B. Heinz[1], Shuaimin Liu[1], Philip Kim[3], James Hone[†,1]

[1] Mechanical Engineering, Columbia University, New York, New York 10027, USA

[2] EHF Key Laboratory of Fundamental Science, School of Electronic Engineering, University of Electronic Science and Technology of China, Chengdu, 611731, China

[3] Department of Physics, Columbia University, New York, New York 10027, USA

[*] These authors contributed equally to this paper. [†]Corresponding email: jh2228@columibia.edu

**We report radio frequency (RF) electrical readout of graphene mechanical resonators. The mechanical motion is actuated and detected directly by using a vector network analyzer (VNA), employing a local gate to minimize parasitic capacitance. A resist-free doubly-clamped sample with resonant frequency ~ 34 MHz, $Q$ factor ~10,000 at 77 K and signal-to-background ratio of over 20 dB, is demonstrated. In addition to being over two orders of magnitude faster than the electrical RF mixing method, this technique paves the way for use of graphene in RF devices such as filters and oscillators.**

RF nanoelectromechanical systems (NEMS) are promising in many areas, such as sensing, studies of fundamental physics, and analog high-frequency signal processing [1-3]. For NEMS, robust transduction of the mechanical motion represents a central challenge; methods employing optical interferometry[4], scanned probe microscopy[5], magnetomotive readout[6], piezoresistive downmixing[7], single electron transistors[8], and quantum point contacts[9] have been successfully demonstrated. In addition, shrinking NEMS down to the smallest scales offers advantages in frequency and responsivity. To this end, carbon nanotubes and graphene (cylinders and single atomic sheets of carbon, respectively) have generated considerable interest due to their intrinsically small size and excellent mechanical properties[10,11]. In addition, these materials possess gate-tunable conductance, which enables fully electrical readout of mechanical motion[12-14]. However, the high device resistances and large parasitic capacitances in these systems have necessitated the use of an electromechanical down-mixing technique. This technique, although quite successful, limits the measurement bandwidth (~ 1 kHz), making these measurements very time consuming, and is not applicable to RF applications (e.g. filters[3] and oscillators[15]). Here, by employing a local gate to reduce the stray capacitance, we demonstrate direct readout (i.e. carried out at the resonant frequency) of graphene RF NEMS resonators.

The device geometry is shown in Fig. 1 (a). The fabrication process starts with dry growth of 100 nm oxide on high resistivity (10 kΩ·cm) silicon wafers, followed by patterning and deposition of a gate electrode. Another 200 nm of oxide, which buries the gate electrodes, is then deposited by plasma-enhanced chemical vapor deposition (PECVD). Source and drain electrodes are then patterned on top of the gate. Finally, a dry etch process is used to open a trench (200 nm deep) between the source and drain contacts, as well as wire-bonding windows for gate electrodes. Following substrate fabrication, graphene is deposited onto the electrodes by mechanical exfoliation[16], and suitable flakes are located by optical microscopy. An advantage of this process is that the graphene is never exposed to polymer resists, and so remains cleaner than graphene that is contacted by lithography. Finally, the device is loaded into a variable-temperature high-vacuum (<$10^{-7}$Torr) probe station.

Fig.1 (b) shows the gate dependence of DC electrical conductance ($G$) of a monolayer graphene flake, as a function of gate DC bias ($V_g$), measured at 77K, and the inset shows the differential conductance. The circuit used to drive and detect the mechanical oscillation of the graphene is shown in Fig. 1 (a): A DC voltage $V_d$ is applied to the drain. An RF drive with amplitude $\tilde{V}_g$ of frequency $\omega/2\pi$ generated by a network analyzer (HP 3577A), and a DC voltage $V_g$ are combined using a bias tee and applied to the gate. The output current is split into DC and RF components using a second bias tee, and the RF component $\tilde{I}$ is fed into the input of the network analyzer. The measured transmission of the device is then $S_{21} = 50\Omega \times \tilde{I}/\tilde{V}_g$, where the factor of 50 Ω is due to the network analyzer input impedance.

Fig. 2 (a) shows the measured |$S_{21}$| of a monolayer graphene device as a function of drive frequency, measured at 77 K. The mechanical resonance appears as a peak ~7 dB in amplitude. This signal-background ratio is comparable with a previously-reported balanced bridge scheme[17]. A significant advantage of direct RF readout is measurement speed: the measurement time for the data shown is 1 s, whereas in our previous studies of graphene NEMS using the mixing technique[14], a comparable dataset typically required ~240 s.

Fig. 2 (b) shows the dependence of dependent |$S_{21}$| on frequency and $V_d$. The resonance appears as a peak for positive $V_d$ and a dip for negative $V_d$, and disappears at $V_d$ = 0. For this device, the

background signal is largely independent of $V_d$. The measured dispersion ($V_g$ dependence) of the resonant frequency is shown in Fig. 2 (c). In this device, the dispersion is negative. In general, both positive dispersion (due to tensioning of the graphene membrane) and negative dispersion (due to a voltage-induced decrease in the effective spring constant) can be observed[14,18]. Fig. 2 (d) shows the measured $|S_{21}|$ with small drive power (-55 dBm), at 77 K. The quality factor ($Q$) reaches ~10,000. The large dip (more than 20 dB) shows that the proposed method has ultra-high sensitivity in detecting the small signals involved in NEMS measurements.

We now turn to analysis of the measured data. In a simple linear oscillator model, the capacitive force from the RF drive should induce an oscillation in the graphene membrane with amplitude

$$\tilde{z} = -\frac{1}{m}\frac{C_g}{z_0}V_g\tilde{V}_g \cdot \frac{1}{\omega_0^2 - \omega^2 + j\omega_0\omega/Q} \tag{1}$$

Here, $m$ is the mass of graphene resonator, $C_g$ and $z_0$ are the capacitance and distance between suspended graphene and gate, respectively, $\omega_0$ and $Q$ are the resonant frequency and quality factor of graphene mechanical resonator, $\omega$ is the frequency of driving force. The RF current consists of four separate terms, of which the first two of are purely capacitive and the last two depend on the active nature of the graphene channel, i.e. its gate-tunable conductance.

$$\tilde{I} = j\omega C_{tot}\tilde{V}_g - j\omega\frac{\tilde{z}}{z_0}C_gV_g + V_d\frac{dG}{dV_g}\tilde{V}_g - V_d\frac{dG}{dV_g}\frac{\tilde{z}}{z_0}V_g \tag{2}$$

The first term is the capacitive background signal, where $C_{tot}$ is the total capacitance between the gate and the drain. The small parasitic capacitance (~ 10 fF)[19] due to the local back gate ensures that this term is small enough to enable detection of the mechanical signal. The second term arises from mechanical oscillation of the graphene, which induces an RF current by modulation of the gate capacitance. This effect is commonly used to detect the motion of microelectromechanical (MEMS) devices, and for MEMS-based signal processing[3]. The third term in Eq. (2) is purely electrical, while the fourth term is mechanical: as the graphene vibrates, the oscillating gate capacitance modulates the conductance.

The data shown in Fig. 2 allow us to discriminate among the four sources of the RF current. First, we note that the third and fourth terms disappear at $V_d = 0$. At this point, the peak from the mechanical resonance disappears as in Fig. 2 (b), indicating that the second term in Eq. (2) is negligible. We do observed very weak resonance while $V_d = 0$ with larger-size samples, suggesting the effect from second term. This observation highlights a key advantage of graphene: capacitive methods used for MEMS shows difficulty when scaling down to the nano regime, but graphene's gate-tunable conductance enables readout even in these ultrasmall devices. Away from the resonant frequency, the variation with $V_d$ allows discrimination between the two background origins, i.e. first and third terms in Eq. (2). In this device, the background is largely independent of $V_d$, indicating that the third term is also negligible (in other devices, a variation of up to ~50% can be observed, indicating comparable contributions from both terms).

From Eq. (2), it is straightforward to understand why the mechanical resonance can appear both as a peak and a dip in. At the resonant frequency, the oscillation amplitude is purely imaginary, so that first and forth terms are in phase. Considering the factors from Eq. (1), the fourth term in Eq. (2) varies as $jV_d V_g^2 \frac{dG}{dV_g}$. For positive $V_g$, this term is negative for negative $V_d$, leading to a dip, and positive for positive $V_d$, leading to a peak, as observed [Fig. 2 (b)]. Likewise, for positive $V_d$, the resonance appears as a dip for negative $V_g$ and a peak for positive $V_g$, due to the behavior of the $dG/dV_g$ term [Fig. 2 (c)]. Finally, (vector) subtraction of the background using both the amplitude and phase of the measured signal permits extraction of the absolute oscillation amplitude, as shown in the inset to Fig. 2 (a). In this device, $\tilde{z}$ reaches ~ 0.22 nm at 33.27 MHz.

In summary, a direct electrical detection method for graphene mechanical resonators has been demonstrated. The device is operated as a resonant channel transistor, in which the effect of the oscillating capacitance is amplified by the transistor action of the graphene channel. Such a structure is reminiscent of both the resonant gate transistor[20], which utilizes a oscillating metallic gate electrode, and the recently-demonstrated resonant body transistor[21], in which the entire transistor structure oscillates. Graphene devices differ from these CMOS-based structures in that they are orders of magnitude lower in mass, and have gate-tunable resonant frequencies. Further scaling down of graphene device size and optimization of the structure for lower parasitic capacitance and higher transconductance may enable readout of graphene NEMS in the GHz range[22] for use in wireless communication and studies of fundamental physics.

We acknowledge support from the Air Force Office of Scientific Research (MURI FA955009-1-0705); NSF under CHE-0117752; and the New York State Office of Science, Technology, and Academic Research (NYSTAR).

1	Craighead, H. G. Nanoelectromechanical systems. *Science* **290**, 1532-1535 (2000).

2      Schwab, K. C. & Roukes, M. L. Putting mechanics into quantum mechanics. *Phys Today* **58**, 36-42 (2005).

3      Nguyen, C. T. C., Katehi, L. P. B. & Rebeiz, G. M. Micromachined devices for wireless communications. *P Ieee* **86**, 1756-1768 (1998).

4      Bunch, J. S. *et al.* Electromechanical resonators from graphene sheets. *Science* **315**, 490-493, doi:Doi 10.1126/Science.1136836 (2007).

5      Garcia-Sanchez, D. *et al.* Imaging mechanical vibrations in suspended graphene sheets. *Nano Lett* **8**, 1399-1403, doi:Doi 10.1021/Nl080201h (2008).

6      Husain, A. *et al.* Nanowire-based very-high-frequency electromechanical resonator. *Appl Phys Lett* **83**, 1240-1242, doi:Doi 10.1063/1.1601311 (2003).

7      Li, M., Tang, H. X. & Roukes, M. L. Ultra-sensitive NEMS-based cantilevers for sensing, scanned probe and very high-frequency applications. *Nat Nanotechnol* **2**, 114-120, doi:Doi 10.1038/Nnano.2006.208 (2007).

8      Knobel, R. G. & Cleland, A. N. Nanometre-scale displacement sensing using a single electron transistor. *Nature* **424**, 291-293, doi:Doi 10.1038/Nature01773 (2003).

9      Poggio, M. *et al.* An off-board quantum point contact as a sensitive detector of cantilever motion. *Nat Phys* **4**, 635-638, doi:Doi 10.1038/Nphys992 (2008).

10     Novoselov, K. S. *et al.* Electric field effect in atomically thin carbon films. *Science* **306**, 666-669, doi:Doi 10.1126/Science.1102896 (2004).

11     Lee, C., Wei, X. D., Kysar, J. W. & Hone, J. Measurement of the elastic properties and intrinsic strength of monolayer graphene. *Science* **321**, 385-388, doi:Doi 10.1126/Science.1157996 (2008).

12     Sazonova, V. *et al.* A tunable carbon nanotube electromechanical oscillator. *Nature* **431**, 284-287, doi:Doi 10.1038/Nature02905 (2004).

13     Huttel, A. K. *et al.* Carbon Nanotubes as Ultrahigh Quality Factor Mechanical Resonators. *Nano Lett* **9**, 2547-2552, doi:Doi 10.1021/Nl900612h (2009).

14     Chen, C. Y. *et al.* Performance of monolayer graphene nanomechanical resonators with electrical readout. *Nat Nanotechnol* **4**, 861-867, doi:Doi 10.1038/Nnano.2009.267 (2009).

15     Feng, X. L., White, C. J., Hajimiri, A. & Roukes, M. L. A self-sustaining ultrahigh-frequency nanoelectromechanical oscillator. *Nat Nanotechnol* **3**, 342-346, doi:Doi 10.1038/Nnano.2008.125 (2008).

16     Blake, P. *et al.* Making graphene visible. *Appl Phys Lett* **91**, -, doi:Artn 063124
Doi 10.1063/1.2768624 (2007).

17     Ekinci, K. L., Yang, Y. T., Huang, X. M. H. & Roukes, M. L. Balanced electronic detection of displacement in nanoelectromechanical systems. *Appl Phys Lett* **81**, 2253-2255, doi:Doi 10.1063/1.1507833 (2002).

18     Singh, V. *et al.* Probing thermal expansion of graphene and modal dispersion at low-temperature using graphene NEMS resonators (vol 21, 165204, 2010). *Nanotechnology* **21**, -, doi:Artn 209801
Doi 10.1088/0957-4484/21/0/209801 (2010).

19     Garg, R. & Bahl, I. J. Characteristics of Coupled Microstriplines. *Ieee T Microw Theory* **27**, 700-705 (1979).

20     Nathanso.Hc, Newell, W. E., Wickstro.Ra & Davis, J. R. Resonant Gate Transistor. *Ieee T*

*Electron Dev* **Ed14**, 117-& (1967).

21	Weinstein, D. & Bhave, S. A. The Resonant Body Transistor. *Nano Lett* **10**, 1234-1237, doi:Doi 10.1021/Nl9037517 (2010).

22	Huang, X. M. H., Zorman, C. A., Mehregany, M. & Roukes, M. L. Nanodevice motion at microwave frequencies. *Nature* **421**, 496-496, doi:Doi 10.1038/421496a (2003).

List of figure captions

Fig. 1 (a) Schematic of local gated graphene mechanical resonator and RF measurement setup. Scale bar: 2 μm. (b) Gate dependence of DC conductivity $G$ at bias current ~ 0.1 μA. Inset shows $dG/dV_g$ over the same range.

Fig. 2 (a) Magnitude of transmission scattering parameter ($|S_{21}|$) at drain bias $V_d$ = -0.1 V. Inset picture is the calculated vibration distance. (b) 3D plot of drain bias $V_d$ dependent $|S_{21}|$ of resonator. (c) Resonant frequency as a function of $V_g$ at $V_d$ = -0.1 V. (d) Linear mechanical response of $|S_{21}|$ biased at $V_d$ = -0.1 V with smaller drive power.

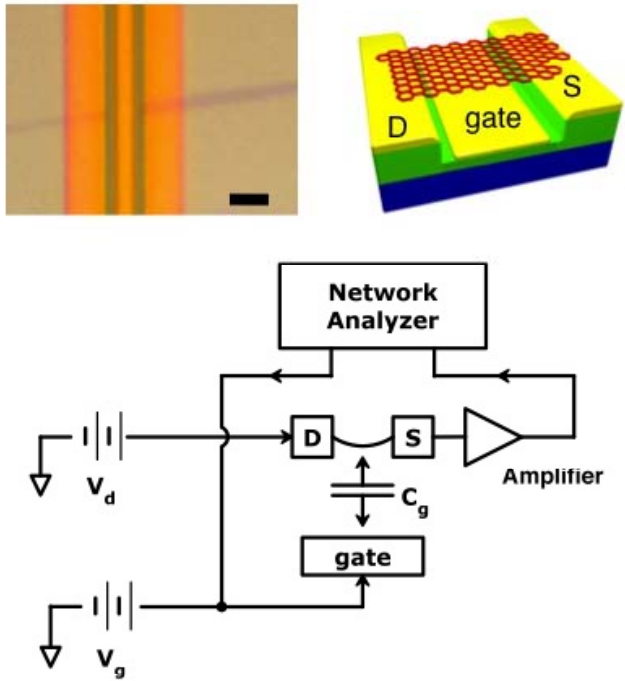
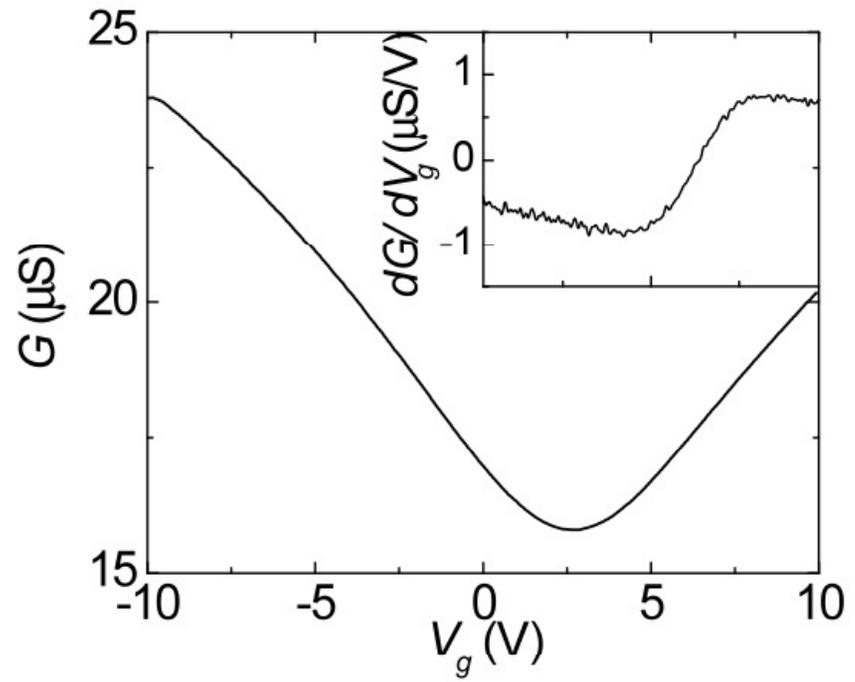

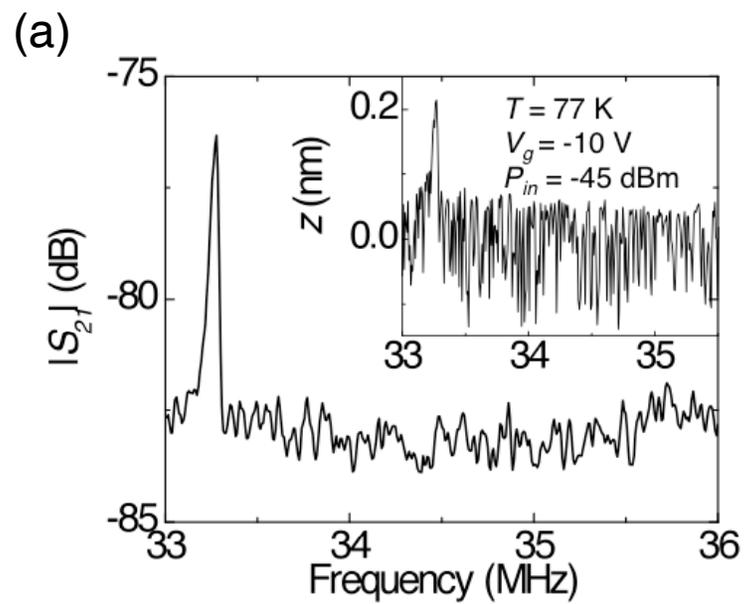
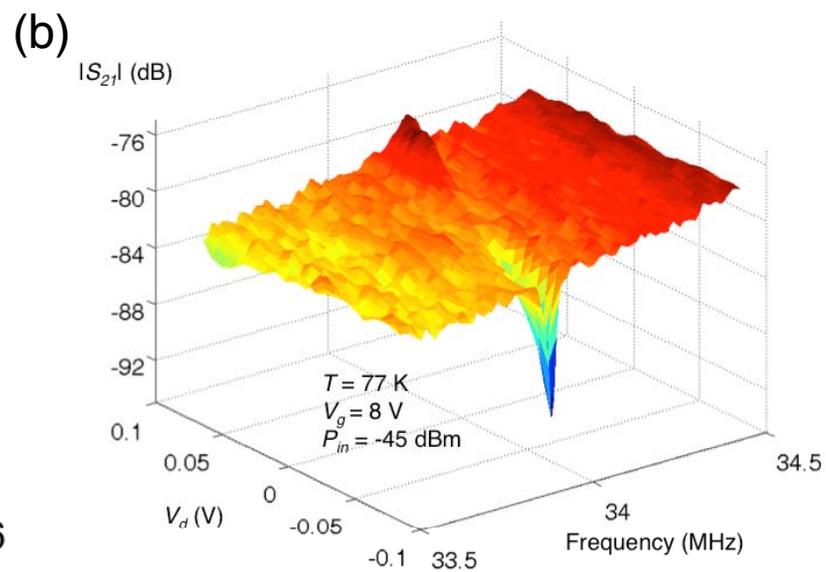
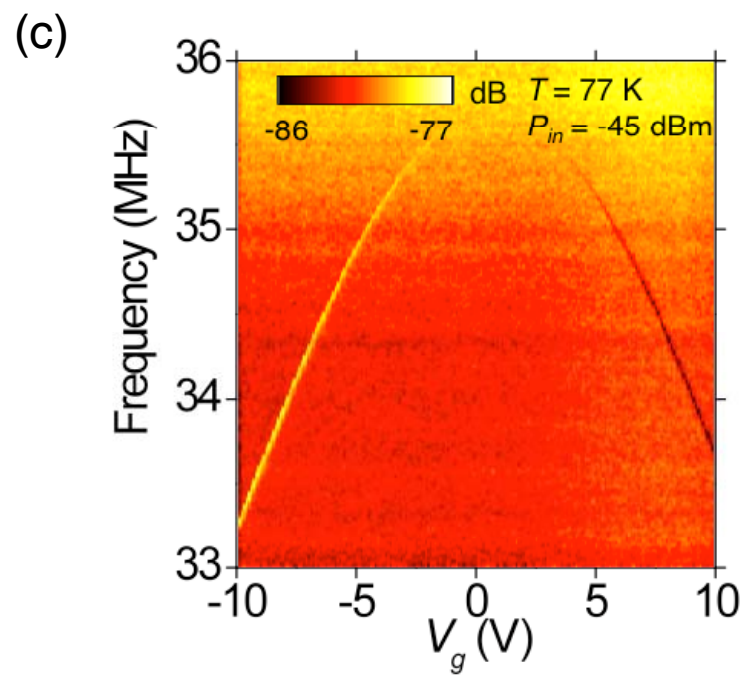
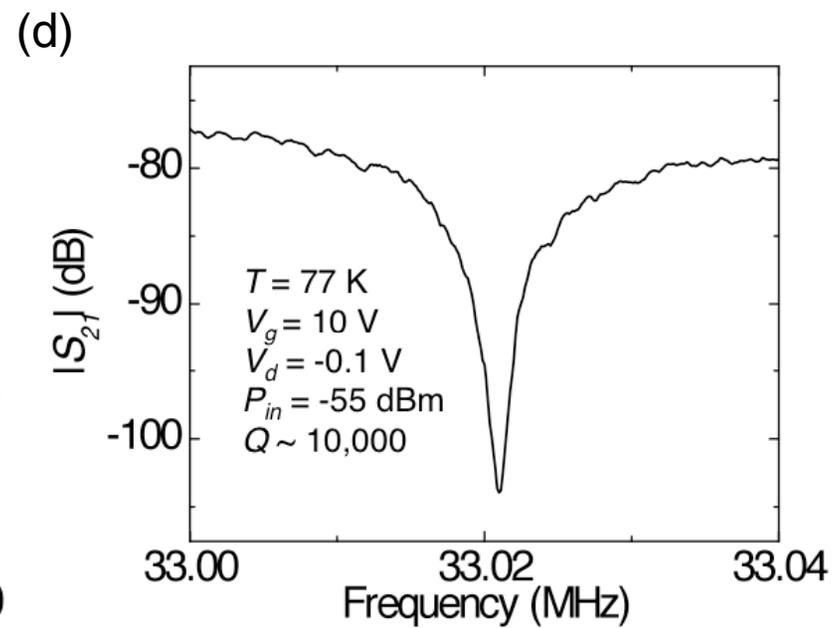